# Optically locked low-noise photonic microwave oscillator

Zexing Zhao[1], Bin Li[1], Kunpeng Jia[1,*], Chenye Qin[1], Xiaofan Zhang[1], Jingru Ji[2], Hua-Ying Liu[1], Xiao-Hui Tian[1], Zhong Yan[3], Xiaoshun Jiang[1], Xinlun Cai[4], Biaobing Jin[1,5], Zhenlin Wang[1], Wei Liang[2,*], Shi-ning Zhu[1], Zhenda Xie[1,*]

[1] *National Laboratory of Solid State Microstructures, School of Electronic Science and Engineering, College of Engineering and Applied Sciences, School of Physics, Research Institute of Superconductor Electronics (RISE) & Key Laboratory of Optoelectronic Devices and Systems with Extreme Performances of MOE, Key Laboratory of Intelligent Optical Sensing and Manipulation, Ministry of Education, and Collaborative Innovation Center of Advanced Microstructures, Nanjing University, Nanjing 210093, China*

[2] *Key Laboratory of Semiconductor Display Materials and Chips, Suzhou Institute of Nano-Tech and Nano-Bionics, Chinese Academy of Sciences, Suzhou 215213, P. R. China*

[3] *Nanzhi Institute of Advanced Optoelectronic Integration, Nanjing 211800, China*

[4] *State Key Laboratory of Optoelectronic Materials and Technologies, School of Electronics and Information Technology, Sun Yat-sen University, Guangzhou, 510275, China*

[5] *Purple Mountain Laboratories, Nanjing 211111, China*

\* E-mail: jiakunpeng@nju.edu.cn; wliang2019@sinano.ac.cn; xiezhenda@nju.edu.cn

# ABSTRACT

The next-generation sensing and communication applications rely on high-frequency microwave generation with low noise. Microwave photonic technology is highly promising; however, its practical applications have so far been limited by the complexity of existing system architectures. Here, we demonstrate an optically locked low-noise photonic microwave oscillator, in which all the optical components are packaged within a small module of 166 mL, and low noise microwave generation is achieved at 10.4 GHz with single-sideband phase noise of −54 dBc/Hz at 10 Hz, −141 dBc/Hz at 10 kHz, and −162 dBc/Hz at 10 MHz offset. Above performance arises from a dual-laser self-injection-locking scheme to a single Fabry–Pérot cavity with high Q exceeding $10^8$, with over 20 dB common-mode noise suppression. The low-noise nature of this reference is coherently transferred to the X-band through a high-performance TFLN electro-optic comb chip, thereby overcoming long-standing barriers in photonic microwave integration to enable truly field-deployable low-noise microwave generation.

# INTRODUCTION

High-performance microwave generation defines capabilities of the electronic communication, radar, and sensing systems. The increasing demand on high communication speed and sensing precision pushes microwave to high operating frequency and low phase noise simultaneously. These stringent requirements pose fundamental challenges for conventional electronic technologies, whose noise performance typically degrades as the operating frequency increases. Microwave photonic techniques leverage the inherently high frequencies of optical oscillators to break this compromise [1-3], transferring the high frequency and low phase noise stability of optical reference into the microwave domain with high fidelity [4-7], albeit at the cost of much greater system complexity in locking compared to electronic methods. Acting as the ultimate "phase anchor" for photonic microwave oscillators, conventional optical references typically rely on large-scale ultra-stable cavities [8]. Their vacuum housings and active thermal isolation systems lead to substantial system size and power consumption, which hinders miniaturization and flexible deployment. This has become a critical bottleneck limiting further practical applications.

To translate high-performance photonic microwave oscillators into practical systems and fully leverage the advantages of photonic integration, optical references need to be implemented in more compact architectures [9-12]. Recently, ultra-high-Q miniature Fabry–Pérot (mini-FP) reference cavities [13-15] offer an absolute optical frequency reference with much lower thermorefractive noise (TRN) and longer photon lifetimes than conventional dielectric resonators [16-19]. However, realizing optical reference–based photonic microwave oscillators (in brief, OR-PMOs) with such ultra-stable FP reference cavity faces several challenges: active locking requires real-time electrical feedback, which leads to bulky and power-hungry architectures, and the FP reference cavity is inherently difficult to integrate into compact, deployable systems. Developing miniaturized optical references with intrinsic frequency stability is expected to pave the way for integrated and deployable low-noise photonic microwave oscillators.

In this work, we demonstrate a compact low-noise photonic microwave oscillator based on an optically locked micro optical reference. Two distributed-feedback (DFB) lasers are simultaneously self-injection locked to a single ultra-high-Q micro Fabry–Pérot (μFP) optical reference cavity with a volume of only 1 mL. This fully optical locking scheme generates dual ultra-narrow-linewidth lasers, with each laser exhibiting a fundamental linewidth of only a few millihertz. By self-injection-locking both lasers to the same μFP cavity, phase-noise fluctuations induced by cavity vibrations and thermal noise are largely suppressed. As a result, more than 20 dB of common-mode noise suppression (CMNS) is achieved. An integrated high-performance thin-film lithium niobate (TFLN) cascaded modulator is used to generate dual electro-optic (EO) frequency combs, serving as the frequency-dividing element in the electro-optical-frequency-division (e-OFD) system. The e-OFD scheme transfers the stabilized optical reference to the X-band, yielding a 10.4 GHz carrier with single-sideband (SSB) phase noise of −54 dBc/Hz at 10 Hz, −141 dBc/Hz at 10 kHz, and −162 dBc/Hz at 10 MHz offset frequency. The oscillator further supports continuous frequency tuning over 350 kHz without mode hopping, highlighting its potential as a compact and deployable platform for high-performance photonic microwave systems.

## RESULTS

### Compact low-noise photonic microwave oscillator system

The architecture of compact low-noise photonic microwave oscillator system based on e-OFD (Fig. 1) comprises a frequency-stabilized micro optical reference, an on-chip frequency divider and a microwave signal generator. The e-OFD scheme employs the frequency divider to phase-lock a tunable electronic oscillator to the optical reference, thereby generating low-noise microwave signals. A key advantage of e-OFD is that all signal processing takes place within the bandwidth of standard electronic equipment, enabling precise tuning to achieve the desired microwave frequency.

At the core of this system is a compact dual self-injection-locked (dSIL) laser that provides the high-performance optical frequency reference. The high-Q external-cavity-based self-injection locking technique enables optical locking between the laser and the reference cavity [20,21] and achieves strong linewidth compression [22-24] without the need for additional electrical feedback loops [25]. In the dSIL configuration, two DFB pump lasers are coupled to the μFP cavity through a micro-optics system. Owing to the ultra-high-Q μFP cavity's frequency selectivity, the fundamental linewidths of the lasers will be significantly compressed by several orders of magnitude, which is crucial for ensuring the frequency stability of optical reference.

The high-performance TFLN modulator generates dual electro-optic (EO) combs that act as frequency dividers. These combs coherently subdivide the stabilized optical reference into equally spaced lines, enabling optical-to-microwave frequency division within the e-OFD system. In recent years, the TFLN platform has demonstrated significant research results in $\chi^{(3)}$ [26-28] and $\chi^{(2)}$ [29,30] based devices. The most distinctive capability of TFLN is the high Pockels coefficient, which enables high-performance EO modulation critical for bridging microwave and optical domains.

TFLN thus has facilitated electro-optic integrated devices including low half-wave voltage ($V_\pi$) modulators [31], EO frequency combs [32,33], and high-speed tunable lasers [34,35], driving progress in optical computing [36], precise detection [37,38], and microwave photonics [39,40]. As the premier electrical-to-optical interface platform, high-performance TFLN is essential for realizing integrated e-OFD technology.

Subsequently, the dSIL lasers are coupled into a high-performance TFLN modulator driven by an external free-running electronic oscillator, generating dual EO combs with identical repetition frequency. The two nearest comb lines in the center of the EO combs produce an intermediate frequency signal $f_{IF} = (f_B - mf_{EO}) - (f_A + nf_{EO})$. The intermediate frequency signal accumulates phase noise jitter scaled by a factor of (n+m) relative to the oscillator [7]. By optimizing microwave parameters, the intermediate frequency signal can be mixed with a stable low-frequency local oscillator. The down-converted error signal then drives a servo feedback loop that divides the original optical reference into the RF domain through an (n+m)-fold frequency division process. Correspondingly, the original phase noise is suppressed by a factor of $(n+m)^2$, enabling the generation of low-noise microwave signals.

**Micro-photonics devices design**

The air-gap high-Q μFP cavity-based dSIL lasers are essential for obtaining a frequency-stabilized optical reference. Critical to dSIL performance, the μFP cavity is fabricated from ultra-low thermal-expansion glass (Zerodur), effectively minimizing thermally induced fluctuations of cavity length. Compared with mini-FP cavities, the μFP achieves comparable Q factors and low thermorefractive noise in a significantly smaller footprint, making its physical scale highly compatible with photonic-chip integration [15]. As depicted in Fig. 2a (with the corresponding optical photograph in Fig. 2d), this air-gap high-Q μFP cavity structure inherently suppresses the TRN-induced frequency drift. Fabrication commenced with ultra-smooth polishing and high-reflection coating of two square plates, which are then assembled to form a hollow FP cavity. The cavity dimensions are 9 mm×9 mm×13 mm, corresponding to a volume of approximately 1 mL. The free spectral range (FSR) near the 1550 nm is 15 GHz and the Q factor is measured to be over $10^8$ (Fig. 2e).

We achieve the optically locked, frequency-stabilized micro optical reference by building a compact dSIL system (Fig. 2b). Two commercial DFB diode lasers centered at 1548.3 nm and 1553.3 nm respectively are employed, with output power of 80 mW and fundamental linewidth of approximately 100 kHz. Each DFB laser is first collimated by a microlens and then directed onto a filter after passing through a silicon-based phase controller. The filter is designed to have a high transmission for one laser and a high reflection for the other, enabling both lasers to be coupled into the µFP cavity through the same optical path while avoiding crosstalk during SIL process. After reflection and beam splitting, the DFB lasers are coupled into µFP cavity. The transmitted light passes through the filter and is spatially aligned with two photodetector chips. These photodetectors independently monitor the SIL states of the two lasers. Meanwhile, the reflected light returns to the respective DFB chips along the original optical path. Through this process, dual ultra-narrow-linewidth self-injection-locked lasers are generated and delivered to an optical fiber coupler, providing a total output power of 3 mW. The linewidth narrowing coefficient η is given by the following formula [41]:

$$\eta \approx \frac{Q_{DFB}^2}{Q_{FP}^2} \frac{1}{16\Gamma_{FP}^2(1+\alpha_g^2)} \quad (1)$$

Where Q factor of the DFB chip $Q_{DFB}$ is 4~6×10$^3$, Q factor of the µFP, $Q_{FP}$ is 2×10$^8$, amplitude-phase coupling coefficient $\alpha_g$ is 2.5, and the reflection coefficient $\Gamma_{FP}$ in our system is 1~3×10$^{-2}$. These parameters are influenced by the manufacturing tolerances of commercial DFB chips, the optical-path coupling efficiency, and variations among the micro-optics components. As a result, the linewidth narrowing factor η falls with the range of 10$^{-7}$ to 10$^{-8}$.

We implement an integrated circuit for conveniently connecting to external leads with the modular packaging (Fig. 2f). The overall volume is approximately 60 mL, and the total weight is 113 g. Notably, the entire micro-optical subsystem occupies only 18 mm×34 mm×9 mm (white dashed frame in Fig. 2f), while the remaining portion of the module consists of a circuit board for external connections and a protective metal enclosure. The system is pumped by external electric circuits to control the DFB lasers, silicon phase shifters, and TEC temperature, thereby enabling both the generation and state control of the dSIL lasers. The SIL status can be monitored through the embedded PD electrical signal output, thereby ensuring robust self-injection locking. This compact, modular packaging enables the deployment of stable optical-reference sources for distributed applications.

For EO combs generation, we employ a TFLN modulator (as shown as Fig. 2c, g) featuring a cascaded Mach-Zehnder modulator (MZM) and three phase modulators (PMs) in series. This architecture significantly reduces RF power requirements per phase modulator while supporting broadband EO comb generation under low-frequency driving conditions. Owing to an efficient optically and electrically folded design (details in Ref. [42]), the TFLN devices achieves a low half-wave voltage (≈1.7

V @ low-frequency for MZM, ≈2 V for PMs). Both optical and RF ports are also packaged within a miniaturized module of ~106 mL for system integration.

**Optically locked frequency-stabilized micro optical reference**

We experimentally validate the performance of packaged ultra-narrow-linewidth dSIL lasers module as a turnkey, high-performance frequency-stabilized micro optical reference (Fig. 3). The characterization through the testing setup as shown in Fig. 3a. Crucially, the dSIL processes within this linear cavity can basically be regarded as independent of each other [43] and are inherently polarization insensitive. This enables independent adjustment of each laser's locking state without cross-interference, thereby significantly enhancing control flexibility.

By applying a triangular-wave modulation to the driven currents of DFB lasers and monitoring the real-time electrical outputs from the embedded photodetectors, we observe clear power transitions associated with the locking and unlocking events of SIL A and SIL B (Figs. 3b and 3c). The measured self-injection locking bandwidths are approximately 1.77 GHz for SIL A and 0.82 GHz for SIL B. We evaluate the turnkey operation of the dSIL lasers module. The optical-reference system reliably initiates and terminates simultaneous self-injection locking of both lasers at 1548.3 nm and 1553.3 nm (Figs. 3d and 3f upper panel). The dSIL lasers exhibit a side-mode suppression ratio (SMSR) of exceeding 50 dB (Fig. 3e) and achieves long-term self-stable maintenance (Fig. 3f lower panel).

We characterize the linewidths of the dSIL lasers using a delay-based heterodyne self-frequency system (details in Methods) [44,45]. Within the detuning range of ~ 300 MHz, the fundamental linewidths of the dSIL lasers remain stable, demonstrating exceptional robustness for optical reference applications (Fig. 3g, h). The results of the frequency noise of the both lasers in the free-running and SIL states are shown in Fig. 3(i). The fundamental linewidths of the SIL laser A and SIL laser B are 9 mHz and 55 mHz, respectively, which are derived from the base of the frequency noise curves [44]. We further use a wavelength meter (Highfinesse WS-U) to assess the long-term frequency stability of the SIL lasers. Thanks to the stable frequency optical locking enabled by air-gap high-Q microcavity and the mechanical rigidity of the packaged module, the central frequencies of the both SIL lasers exhibit a drift of only a few megahertz over an 8-hour duration. The relatively intensity noise (RIN) of the dSIL lasers (by using Rohde & Schwarz FSV30) are demonstrated in Fig. 3k. Both RIN curves fall below −135 dB/Hz beyond 10 kHz offset, confirming the excellent power stability of dSIL lasers as optical reference.

Notably, by simultaneously locking two lasers to a common μFP in a compact configuration, the scheme naturally rejects common-mode noise arising from cavity vibrations and thermal fluctuations in the dSIL lasers system. As a result, the heterodyne beat note between the two SIL lasers directly reveals the effectiveness of CMNS. The phase noise of the dual SIL lasers' beat note (as shown in Fig. 4c) is measured via two wavelength delayed self-heterodyne interferometer (TWDI, measured details in the Methods). Compared with the individual SIL laser phase noise (grey curves), the CMNS of the dual SIL lasers' beat note reached approximately 20 dB (pink curve).

**Low-noise tunable microwave signal generation**

Figure 4a illustrates the microwave-optical oscillation circuit of e-OFD photonic microwave generation system. The dSIL lasers module first generates dSIL lasers at 1548.3 nm and 1555.3 nm, corresponding to an optical bandwidth of 625 GHz for the OFD. The dSIL lasers are then optically amplified and coupled into the packaged, integrated TFLN cascade modulator. An external dielectric resonator oscillator (DRO) RF source provides a 10.4 GHz microwave signal. The microwave signal is transmitted into the TFLN cascade modulator through the RF link (with a total loaded RF power of approximately 35 dBm), generating dual EO combs (Fig. 4b). The theoretical noise-suppression level achieved through frequency division, given by $\log_{10}(\frac{625\ GHz}{10.4\ GHz})^2 = 35.5$ dB, is expected. We use an optical filter to select adjacent comb line from the dual EO combs, which are detected by a photodetector to produce an intermediate frequency ($f_{IF} \approx 100$ MHz). This signal is then mixed with a 100 MHz local oscillator (LO) to generate an error signal, which is fed into servo controller and subsequently applied as feedback to the DRO. Once the $f_{IF}$ signal is locked, the power spectral density of the microwave phase noise is suppressed by the square of the frequency-division factor over the frequency range within the servo's control bandwidth.

We use phase noise analyzer to characterize the phase noise of the output microwave signals in the both free-running and locked states (Fig. 4c). The SSB phase noise of the locking 10.4 GHz signal from the compact low-noise photonic microwave oscillation system achieves −54 dBc/Hz at 10 Hz offset, −77 dBc/Hz at 100 Hz, −108 dBc/Hz at 1 kHz, −141 dBc/Hz at 10 kHz, −146 dBc/Hz at 100 kHz, −152 dBc/Hz at 1 MHz, −162 dBc/Hz at 10 MHz. Dividing the beat note phase noise by the frequency division factor of 35.5 dB yields excellent agreement with the measured microwave phase noise, with minor deviations above 10 kHz attributed to PNA measurement limitations and servo-feedback bandwidth constraints. Nevertheless, the optically locked frequency-stabilized optical reference enables exceptional phase noise performance beyond 10 kHz offset. The fractional frequency stability of the output microwave signal is presented in Fig. 4e.

We further investigate the frequency tunability of the low-noise photonic microwave oscillator (Figure 5). The output microwave frequency is equal to $\frac{f_B - f_A - f_{IF}}{m+n}$, where $f_{IF}$ is locked to the local oscillator reference. By adjusting the local oscillator frequency, the output microwave frequency can be precisely tuned. A triangular-wave modulation with a period of 100 Hz is applied to the local oscillator, producing a 21 MHz frequency deviation in the reference. The locking system tracks this modulation, resulting in a corresponding microwave frequency deviation of $\frac{21 \text{ MHz}}{60} = 350 \text{ kHz}$ in the low-noise microwave output. Notably, the tuning bandwidth is only limited by the tuning range of the local oscillator.

**CONCLUSION AND OUTLOOK**

While OR-PMOs have demonstrated remarkable low-noise microwave performance, efforts toward miniaturization have historically been limited by the complexity of generating a frequency-stabilized optical reference. Here, we present a compact e-OFD system based on dSIL lasers optically locked to a high-Q μFP reference cavity. The resulting frequency-stabilized optical reference is subsequently converted into dual EO combs via a monolithic TFLN cascaded modulator, enabling low-noise microwave generation with 35.5 dB frequency-division noise suppression and single-sideband phase noise of −54 dBc/Hz at 10 Hz, −141 dBc/Hz at 10 kHz, and −162 dBc/Hz at 10 MHz offset. All of this is realized within a compact, fully integrated module that effectively combines miniaturization with deployability.

In subsequent work, the Q factor of the μFP cavity can be further optimizing by adjusting its end-face reflectivity, reducing roughness, or implementation of vacuum-gap design [13-15]. Meanwhile, the performance of the EO-based division system can be further improved through two complementary strategies: first, the output microwave frequency can be easily scaled to other microwave bands by choosing different DRO, as the repetition rate of EO frequency comb can be continuously tuned [42]; and second, by broadening the EO comb spectrum and increasing the spectral separation between the reference lasers to enhance the frequency-division ratio [46,47]. With these advancements, better noise performance can be anticipated. On the other hand, the current modules for the optical reference and frequency divider are still limited in size by the electrical packaging constraints. Further integration of the micro-optical subsystem with the TFLN chip through advanced micro-photonic coupling techniques will substantially reduce the overall footprint [15], with remaining challenges primarily associated with efficient mode-field matching, low-loss chip–to–chip coupling, and heterogeneous integration across dissimilar material platforms. Overall, our architecture not only significantly reduces the size, weight, power, and cost (SWaP-C) of the complete OFD system but also paves the way for deployment in practical applications, including precision timing, sensing, and communication on next-generation mobile platforms [48,49].

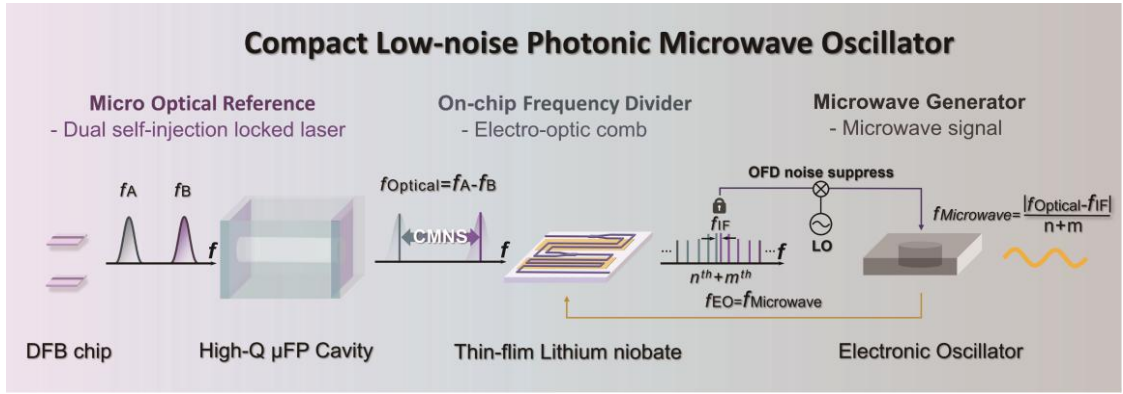

**Figure 1.** Concept of compact low-noise photonic microwave oscillator based on high-Q µ-Fabry-Pérot (µFP) reference cavity. Photonic microwave oscillator system is mainly composed of three parts: i) Micro optical reference. A miniature micro-optics system was constructed to produce dual self-injection locked laser with high common-mode noise suppression (CMNS) serves as frequency-stabilized micro optical references; ii) On-chip frequency divider. High-performance thin-film lithium niobate cascade modulator generates electro-optic combs. New intermediate frequency signals are produced at the two nearest dual EO comb with amplified phase noise information; iii) Microwave generator. The tunable electronic oscillator generate microwave signals used to drive the modulator. The error signal generated by the intermediate frequency signal and the local oscillator (LO) signal is fed back to the oscillator through servo control to form a microwave oscillation loop, generating a low-noise microwave signal.

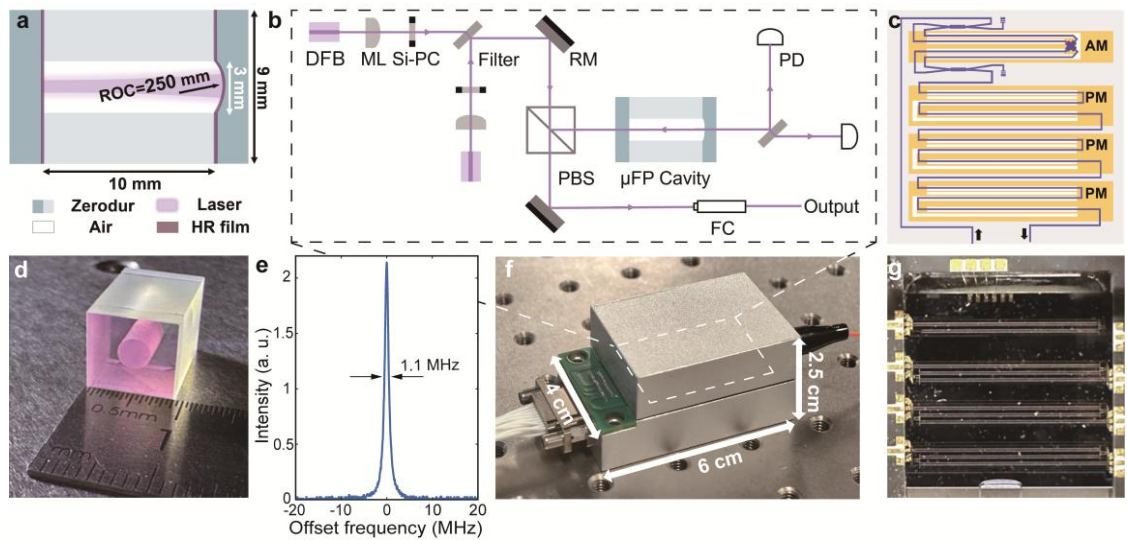

**Figure 2.** (a) Schematic of the high-Q μFP cavity. ROC, radius of curvature. (b) Schematic of the micro-optics system for generating dSIL lasers based on μFP cavity. ML, micro-lens. Si-PC, Silicon-based phase controller. RM, reflect mirror. PBS, polarization beam splitter. PD, photodetector. FC, fiber collimator. (c) TFLN cascade modulator architecture. (d) Optical photograph of the μFP cavity. (e) Q-factor characterization at 1550 nm. (f) Optical photograph of the dSIL lasers packaging module; white dashed area: micro-optics system in (b). (g) Optical photograph of the integrated TFLN modulator.

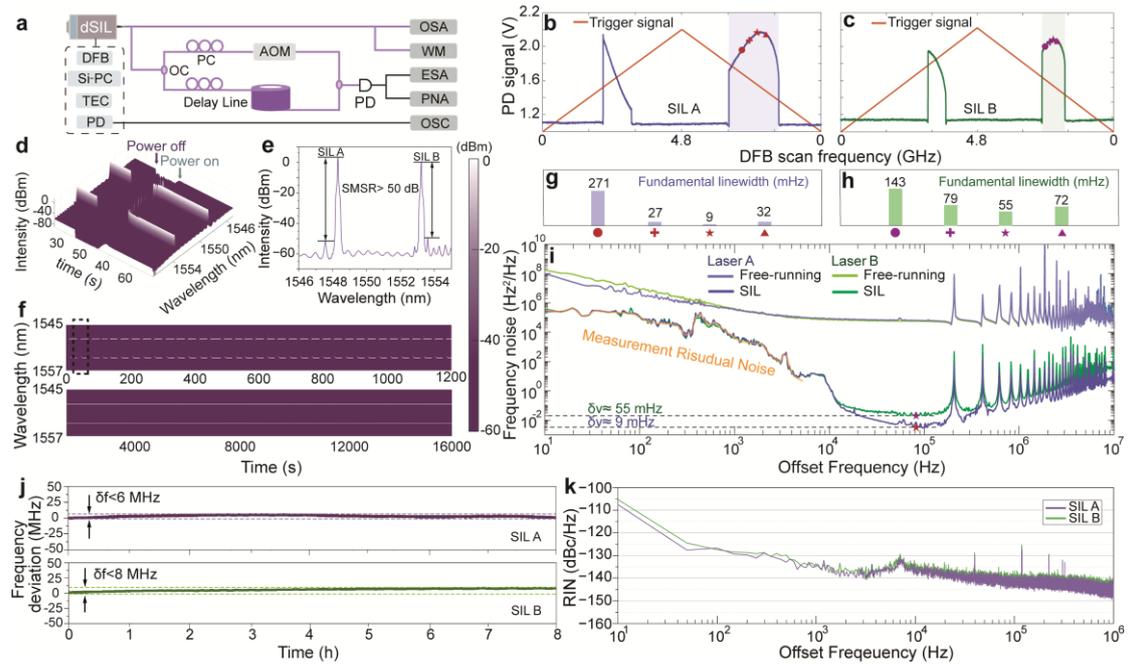

**Figure 3.** (a) The dSIL lasers test setup. TEC, Thermo-Electric-Cooler. OC, optical coupler. PC, polarization controller. AOM, acousto-optic modulator. OSA, optical spectrum analyzer. WM, wavelength meter. ESA, electrical spectrum analyzer. PNA, phase noise analyzer. OSC, oscilloscope. (b and c) The detected PDs' monitor signals of dSIL lasers module when scanning the driven current. (d) The spectral evolution of the turnkey dSIL lasers (the black dashed box in the (f)). (e) Optical spectra of dSIL lasers. (f) The spectrum evolution of dSIL lasers in consecutive turnkey operation (upper) and in long-term free-running state (lower). (g and h) Fundamental linewidths variation (locations marked in the (b) and (c)). (i) The frequency noise characterization of the lasers in free-running and SIL states (resonance center). Dashed horizontal indicates noise floor for fundamental linewidth calculation. (j) Long-term frequency drift measured by WM. (k) The measured RIN of dSIL lasers.

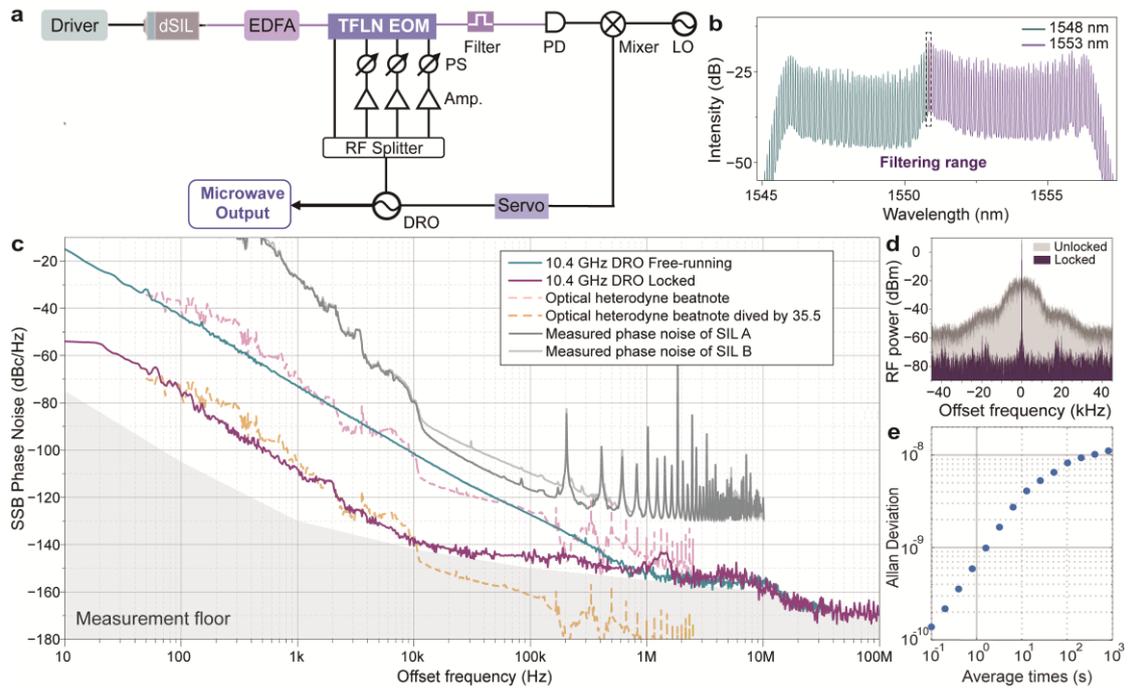

**Figure 4.** (a) Schematic of compact e-OFD system. EDFA, Erbium-doped fiber amplifier. LO, local oscillator. PS, phase shifter. DRO, dielectric resonator oscillator. (b) Dual EO combs spectrum. Dotted line area indicates comb line filtered for photodetection to generate intermediate frequency signal. (c) SSB phase noise of the 10.4 GHz microwave signal in both the free-running and locked states. The phase noise of the optical reference is obtained using optical heterodyne detection, and the corresponding phase noise after e-OFD is calculated. The phase noise of the original dual free-running lasers was plotted for comparison. (d) The electrical spectra of intermediate frequency signals in locked and unlocked states. (e) Allan deviation of the locked 10.4 GHz DRO.

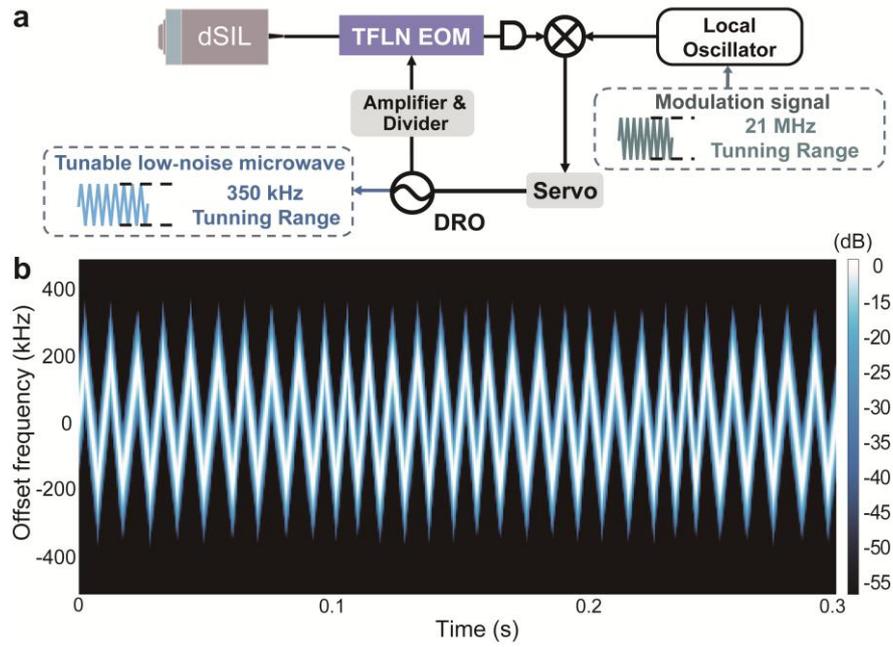

**Figure 5.** (a) Schematic of the frequency tuning test setup for locked 10.4 GHz DRO. A linear frequency modulation signal is applied to the local oscillator, enabling the tuning of the output microwave frequency. (b) Measured offset frequency of the output microwave, characterized by a real-time spectrum analyzer (RBW 1 kHz).

# METHOD

## Heterodyne beat frequency linewidth test system and dual-wavelength beat note noise measured method

We characterize the linewidths and the phase noise of the beat note using a delay-based self-heterodyne interferometer system. The dual SIL lasers are respectively injected into the self-heterodyne interferometer system. The lasers are evenly divided into two paths: one path passes through an acousto-optic frequency shifter with 40 MHz modulation frequency, while the other path passes through 1 km optical fiber to introduce an optical delay $\tau$. After combining the two optical signals, a photodetector is used to detect the beat signal, converting the optical phase information to the RF band. We use the PNA to measure the SSB frequency noise $S_M(f)$ of the RF signal. Combined with following formula (2) [44,45,50], we calculate the frequency noise $S_\phi(f)$ of two lasers in self-injection locked state and free-running. The fundamental linewidths of the lasers could be calculated by $\pi$ times of the base noise.

$$S_\phi(f) = \frac{S_M(f)}{4\sin^2(\pi f \tau)} \qquad (2)$$

Since the measured noise represents the superposition of laser noise and fiber noise, the low offset frequencies region (<5 kHz) of the test results is influenced by the residual noise from the fiber test system. To account for this, we use an additional frequency-stabilized laser to measure the low offset frequencies noise of the test system (yellow curve in Fig. 3i). Nevertheless, due to the sufficient length of our delay fiber, this low offset frequencies does not affect the extraction of noise information in the high-frequency region (>10 kHz) [15].

We measure the phase noise of the beat note frequency between the dSIL lasers by using the two-wavelength delayed self-heterodyne interferometer (TWDI) method [9,51-53]. When dSIL lasers simultaneously input into the self-heterodyne system, two self-heterodyne beat note signals (each with frequency shift and delay) are generated simultaneously. The key aspect of this method lies in the mixing these two beat-note signals. The voltage noise density spectrum of the resulting DC signal contains the phase noise information of the original dual lasers beat note. This phase noise can be extracted through Hilbert variation and calculated using formula (2) to obtain the phase noise of the dSIL lasers beat note.

**ACKNOWLEDGEMENTS:** We thank Jiachuan Yang of Peking University for fruitful discussion.

**FUNDING:** This work was supported by the National Key Research and Development Program of China (2022YFA1205100); National Natural Science Foundation of China (62288101, 62293523, 62293520, 12304421, 12341403, 92463304, 92463308, 623B2047); Zhangjiang Laboratory (ZJSP21A001); Guangdong Major Project of Basic and Applied Basic Research (2020B0301030009); the Natural Science Foundation of Jiangsu Province (BK20230770, BK20232033); Key project of Basic Research Program of Jiangsu Province (BK20253015); Major Project of Scientific and Technological Innovation 2030 (2023ZD0301500).

**AUTHOR CONTRIBUTIONS:** Z.X, K.J., W.L., and Z.Z. conceived the original idea and designed the experiment. Z.Z., K.J., and W.L. prepared the μFP cavity and the packaged optical reference module. X.C., X.J., and Z.Y. provided experimental equipment. Z.Z. and B.L. conducted the experiments and data analysis. Z.Z., K.J., W.L., and Z.X. wrote the manuscript. C.Q., X.Z., J.J., H.-Y. L., X.-H. T., B.J., and Z.W. provided valuable feedback and comments. Z.X, K.J., W.L., and S.Z. supervised the whole work. All authors contributed to the manuscript preparation.

**Conflicts of interest statement.** None declared.

## REFERENCES

1   Capmany J & Novak D. Microwave photonics combines two worlds. *Nat Photon* 2007; **1**: 319–330.
2   Ghelfi P, Laghezza F, Scotti F *et al.* A fully photonics-based coherent radar system. *Nature* 2014; **507**: 341–345.
3   Xie X, Bouchand R, Nicolodi D *et al.* Photonic microwave signals with zeptosecond-level absolute timing noise. *Nat Photon* 2016; **11**: 44–47.
4   Yao J. Microwave Photonics. *J Lightwave Technol* 2009; **27**: 314–335.
5   Hao T, Liu Y, Tang J *et al.* Recent advances in optoelectronic oscillators. *Adv Photon* 2020; **2**: 044001.
6   Fortier T M, Kirchner M S, Quinlan F *et al.* Generation of ultrastable microwaves via optical frequency division. *Nat Photon* 2011; **5**: 425–429.
7   Li J, Yi X, Lee H *et al.* Electro-optical frequency division and stable microwave synthesis. *Science* 2014; **345**: 309–313.
8   Ludlow A D, Boyd M M, Ye J *et al.* Optical atomic clocks. *Rev Mod Phys* 2015;

**87**: 637–701.

9   Tetsumoto T, Nagatsuma T, Fermann M E *et al.* Optically referenced 300 GHz millimetre-wave oscillator. *Nat Photon* 2021; **15**: 516–522.

10  Groman W, Kudelin I, Lind A *et al.* Photonic millimeter-wave generation beyond the cavity thermal limit. *Optica* 2024; **11**: 1583–1587.

11  Niu R, Hua T-P, Shen Z *et al.* Ultralow-Noise K-Band Soliton Microwave Oscillator Using Optical Frequency Division. *ACS Photon* 2024; **11**: 1412–1418.

12  Jin X, Xie Z, Zhang X *et al.* Microresonator-referenced soliton microcombs with zeptosecond-level timing noise. *Nat Photon* 2025; **19**: 630–636.

13  Kelleher M L, McLemore C A, Lee D *et al.* Compact, portable, thermal-noise-limited optical cavity with low acceleration sensitivity. *Opt Express* 2023; **31**: 11954–11965.

14  Liu Y, Jin N, Lee D *et al.* Ultrastable vacuum-gap Fabry–Perot cavities operated in air. *Optica* 2024; **11**: 1205–1211.

15  Cheng H, Xiang C, Jin N *et al.* Harnessing micro-Fabry–Pérot reference cavities in photonic integrated circuits. *Nat Photon* 2025; **19**: 992–998

16  Kudelin I, Groman W, Ji Q-X *et al.* Photonic chip-based low-noise microwave oscillator. *Nature* 2024; **627**: 534–539.

17  Kudelin I, Shirmohammadi P, Groman W *et al.* An optoelectronic microwave synthesizer with frequency tunability and low phase noise. *Nat Electron* 2024; **7**: 1170–1175.

18  He B, Yang J, Meng F *et al.* Highly coherent two-color laser and its application for low-noise microwave generation. *Nat Commun* 2025; **16**: 4034.

19  Ji Q-X, Zhang W, Savchenkov A *et al.* Dispersive-wave-agile optical frequency division. *Nat Photon* 2025; **19**: 624–629.

20  Shen B, Chang L, Liu J *et al.* Integrated turnkey soliton microcombs. *Nature* 2020; **582**: 365–369.

21  Lihachev G, Bancora A, Snigirev V *et al.* Frequency agile photonic integrated external cavity laser. *APL Photon* 2024; **9**: 126102.

22  Guo J, McLemore C A, Xiang C *et al.* Chip-based laser with 1-hertz integrated linewidth. *Sci Adv* 2022; **8**: eabp9006.

23  Jin W, Yang Q-F, Chang L *et al.* Hertz-linewidth semiconductor lasers using CMOS-ready ultra-high-Q microresonators. *Nat Photon* 2021; **15**: 346–353.

24  Qin C, Jia K, Zhao Z *et al.* Compact low-noise dual microcombs for high-precision ranging and spectroscopy applications. *Optica* 2025; **12**.

25  Kondratiev N M, Lobanov V E, Shitikov A E *et al.* Recent advances in laser self-injection locking to high-Q microresonators. *Front Phys* 2023; **18**: 21305.

26  He Y, Yang Q-F, Ling J *et al.* Self-starting bi-chromatic LiNbO3 soliton microcomb. *Optica* 2019; **6**: 1138–1144.

27  Song Y, Hu Y, Zhu X *et al.* Octave-spanning Kerr soliton frequency combs in dispersion- and dissipation-engineered lithium niobate microresonators. *Light Sci Appl* 2024; **13**: 225.

28  Nie B, Lv X, Yang C *et al.* Soliton microcombs in X-cut LiNbO3 microresonators. *eLight* 2025; **5**: 15.

29  Shi X, Mohanraj S S, Dhyani V *et al.* Efficient photon-pair generation in layer-poled lithium niobate nanophotonic waveguides. *Light Sci Appl* 2024; **13**: 282.

30  Chen P K, Briggs I, Cui C *et al.* Adapted poling to break the nonlinear efficiency limit in nanophotonic lithium niobate waveguides. *Nat Nanotechnol* 2024; **19**: 44–50.

31  Wang C, Zhang M, Chen X *et al.* Integrated lithium niobate electro-optic modulators operating at CMOS-compatible voltages. *Nature* 2018; **562**: 101–104.

32  Zhang M, Buscaino B, Wang C *et al.* Broadband electro-optic frequency comb generation in a lithium niobate microring resonator. *Nature* 2019; **568**: 373–377.

33  Hu Y, Yu M, Buscaino B *et al.* High-efficiency and broadband on-chip electro-optic frequency comb generators. *Nat Photon* 2022; **16**: 679–685.

34  Siddharth A, Bianconi S, Wang R N *et al.* Ultrafast tunable photonic-integrated extended-DBR Pockels laser. *Nat Photon* 2025; **19**: 709–717.

35  Xue S, Li M, Lopez-Rios R *et al.* Pockels laser directly driving ultrafast optical metrology. *Light Sci Appl* 2025; **14**: 209.

36  Zhang S, Liu Z, Jiang X *et al.* Thin-film lithium niobate photonic circuit for ray tracing acceleration. *Nat Commun* 2025; **16**: 5938.

37  Qi Y, Jia X, Wang J *et al.* 1.79-GHz acquisition rate absolute distance measurement with lithium niobate electro-optic comb. *Nat Commun* 2025; **16**: 2889.

38  Zhu S, Zhang Y, Feng J *et al.* Integrated lithium niobate photonic millimetre-wave radar. *Nat Photon* 2025; **19**: 204–211.

39  Feng H, Ge T, Guo X *et al.* Integrated lithium niobate microwave photonic processing engine. *Nature* 2024; **627**: 80–87.

40  Ye K, Feng H, te Morsche R *et al.* Integrated Brillouin photonics in thin-film lithium niobate. *Sci Adv* 2025; **11**: eadv4022.

41  Liang W, Ilchenko V S, Eliyahu D *et al.* Ultralow noise miniature external cavity semiconductor laser. *Nat Commun* 2015; **6**: ncomms8371.

42  Wang J, Wang Q, Xu M *et al.* Highly tunable flat-top thin-film lithium niobate electro-optic frequency comb generator with 148 comb lines. *Opt Express* 2025; **33**: 23431–23439.

43  Chermoshentsev D A, Shitikov A E, Lonshakov E A *et al.* Dual-laser self-injection locking to an integrated microresonator. *Opt Express* 2022; **30**: 17094–17105.

44  Rubiola E, Salik E, Huang S *et al.* Photonic-delay technique for phase-noise measurement of microwave oscillators. *J Opt Soc Am B* 2005; **22**: 987–997.

45  Camatel S & Ferrero V. Narrow Linewidth CW Laser Phase Noise Characterization Methods for Coherent Transmission System Applications. *J Lightwave Technol* 2008; **26**: 3048–3055.

46  Weng W, Anderson M H, Siddharth A *et al.* Coherent terahertz-to-microwave link using electro-optic-modulated Turing rolls. *Physical Review A* 2024; **164**: 023511.

47  Loh W, Gray D, Irion R *et al.* Ultralow noise microwave synthesis via difference

|    | frequency division of a Brillouin resonator. *Optica* 2024; **11**: 492–497. |
|----|---|
| 48 | Hilton A P, Offer R F, Klantsataya E *et al.* Demonstration of a mobile optical clock ensemble at sea. *Nat Commun* 2025; **16**: 6063. |
| 49 | Liu J. Building atomic clock lasers with integrated photonics. *Nat Photon* 2025; **19**: 226–227. |
| 50 | Laurent P, Clairon A & Breant C. Frequency noise analysis of optically self-locked diode lasers. *IEEE J Quantum Electron* 1989; **25**: 1131–1142. |
| 51 | Kuse N & Fermann M E. Electro-optic comb based real time ultra-high sensitivity phase noise measurement system for high frequency microwaves. *Sci Rep* 2017; **7**: 2847. |
| 52 | Kuse N & Fermann M E. A photonic frequency discriminator based on a two wavelength delayed self-heterodyne interferometer for low phase noise tunable micro/mm wave synthesis. *Sci Rep* 2018; **8**: 13719. |
| 53 | Lao C, Jin X, Chang L *et al.* Quantum decoherence of dark pulses in optical microresonators. *Nat Commun* 2023; **14**: 1802. |